\documentclass[a4paper,orivec,runningheads]{llncs}
\usepackage{makeidx}  
\usepackage{graphicx}
\usepackage{subfigure}
\usepackage{amsmath}
\usepackage{amssymb}
\usepackage{verbatim}
\usepackage{algorithm}
\usepackage{algorithmic}
\usepackage{booktabs}
\usepackage{multirow}
\usepackage{pbox} 

\begin{document}

\pagenumbering{arabic}

%

%
%
\title{Automatic Segmentation of  Vestibular Schwannoma from T2-Weighted MRI by Deep Spatial Attention with Hardness-Weighted Loss}

\titlerunning{Automatic Segmentation of  Vestibular Schwannoma by Deep Attention}  
%

\author{Guotai Wang\inst{1,2,3} 
\and Jonathan Shapey\inst{2,3,7}
\and Wenqi Li \inst{4}
\and Reuben Dorent\inst{2} 
\and Alex Demitriadis\inst{5}
\and Sotirios Bisdas\inst{6}
\and Ian Paddick \inst{5}
\and Robert Bradford \inst{5,7}
\and S\'ebastien Ourselin\inst{2}
\and Tom Vercauteren\inst{2}}

\authorrunning{G. Wang, et al.} 

\institute{
$^1$ School of Mechanical and Electrical Engineering, University of Electronic Science and Technology of China, Chengdu, China \\
$^2$School of Biomedical Engineering and Imaging Sciences, King's College London, London, UK\\
$^3$Wellcome/EPSRC Centre for Interventional and Surgical Sciences, University College London, London, UK \\
$^4$ NVIDIA, Cambridge, UK \\
$^5$Queen Square Radiosurgery Centre (Gamma Knife), National Hospital for Neurology and Neurosurgery, London, UK\\
$^6$Neuroimaging Analysis Centre, Queen Square, London, UK \\
$^7$Department of Neurosurgery, National Hospital for Neurology and Neurosurgery, London, UK \\
\email{guotai.wang@uestc.edu.cn}\\
}

\maketitle              

\begin{abstract}
Automatic segmentation of vestibular schwannoma (VS) tumors from  magnetic resonance imaging (MRI) would facilitate efficient and accurate volume measurement to guide patient management and improve clinical workflow. The accuracy and robustness is challenged by low contrast, small target region and low through-plane resolution. We introduce a 2.5D convolutional neural network (CNN) able to exploit the different in-plane and through-plane resolutions encountered in standard of care imaging protocols. We use an attention module to enable the CNN to focus on the small target and propose a supervision on the learning of attention maps for more accurate segmentation. Additionally, we propose a hardness-weighted Dice loss function that gives higher weights to harder voxels to boost the training of CNNs. Experiments with ablation studies on the VS tumor segmentation task show that: 1) the proposed 2.5D CNN outperforms its 2D and 3D counterparts, 2) our supervised attention mechanism outperforms unsupervised attention, 3) the voxel-level hardness-weighted Dice loss can improve the performance of CNNs. Our  method achieved an average Dice score and ASSD of 0.87 and 0.43~mm respectively. This will facilitate patient management decisions in clinical practice.

\end{abstract}

\section{Introduction}
Vestibular schwannoma (VS) is a benign tumor arising from one of the balance nerves connecting the brain and inner ear. The incidence of VS has risen significantly in recent years and is now estimated to be between 14 and 20 cases per million per year~\cite{MOFFAT1995}. High-quality magnetic resonance imaging (MRI) is required for diagnosis and expectant management with serial imaging is usually advised for smaller tumors.  Current MR protocols include contrast-enhanced T1-weighted (ceT1) and high-resolution T2-weighted (hrT2) images, but there is increasing concern about the potentially harmful cumulative side-effects of gadolinium contrast agents. 
Accurate measurement of VS tumor volume from MRI is desirable for growth detection and guiding management of the tumor. However, current clinical practice relies on labor-intensive manual segmentation.

\begin{figure*}[t]
	\centering 
	\includegraphics[width=1.0\textwidth]{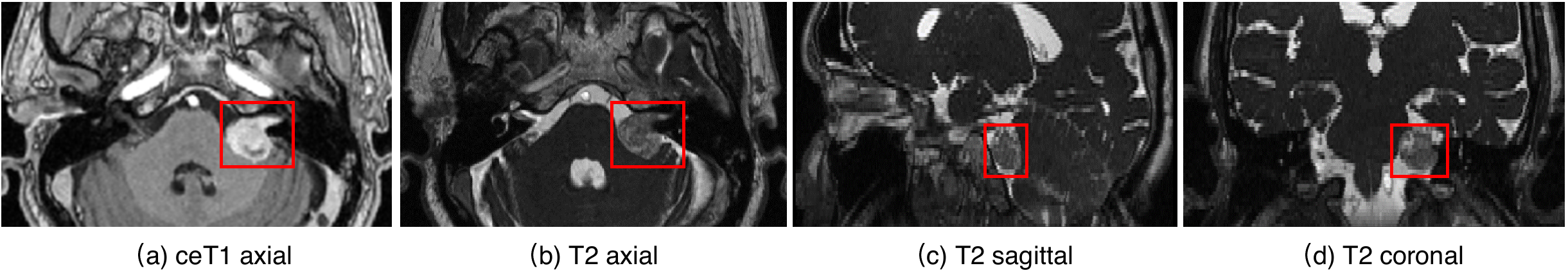}
	\caption{
		An example of VS tumor. (a): contrast-enhanced T1-weighted MRI. (b)-(d): T2-weighted MRI. Note the small target region, low contrast in T2,  and low resolution in sagittal and coronal views.}
	\label{fig:vs_example}
\end{figure*}

This paper aims for automatic segmentation of the VS tumor from high-resolution T2-weighted MRI.  This will improve clinical workflow and enable patients to undergo surveillance imaging without the need for gadolinium contrast, thus improving patient safety. However, this task is challenging due to several reasons. First, T2 images have a relatively low contrast and the exact boundary of the tumor is hard to detect. Second, the VS tumor is a relatively small structure with large shape variations in the whole brain image. Additionally, the image is often acquired with low through-plane resolution, as shown in Fig.~\ref{fig:vs_example}.

In the literature, a Bayesian model was proposed for automatic VS tumor segmentation from ceT1 MRI~\cite{Vokurka2002}, but it can hardly be applied to T2 images with much lower contrast. Semi-automated tools for this task suffer from inter-operator variations~\cite{Tysome2018}. In recent years, convolutional neural networks (CNNs) have achieved state-of-the-art performance for many  segmentation tasks~\cite{Abdulkadir2016,Gibson2018,Hefny2015a}. However, most of them are proposed to segment images with isotropic resolution, and are not readily applicable to our VS images with high in-plane resolution and low through-plane resolution. To segment small structures from large image contexts, 
Yu et al.~\cite{Yu2017} used a coarse-to-fine approach with recurrent saliency transformation. Oktay et al.~\cite{Oktay2018} learned an attention map to enable the CNN to focus on target structures. 
However, the attention map was not learned with explicit supervision during training, and may not be well-aligned with the target region, which can limit the segmentation accuracy. Therefore, we hypothesise that end-to-end supervision on the learning of attention map will lead to better results.
Complementary approaches to deal with small structures include the use of adapted loss functions such as
Dice loss~\cite{Milletari2016} and generalized Dice loss~\cite{Sudre2017}. They can mitigate the class imbalance between foreground and background by image-level weighting during training. Considering the fact that some voxels are harder than the others to learn during training, we propose a voxel-level hardness-weighted Dice loss function to further improve the segmentation accuracy.

The contribution of this paper is three-fold. First, to the best of our knowledge, this is the first work on automatic VS tumor segmentation using deep learning. We propose a 2.5D CNN combining 2D and 3D convolutions to deal with the low through-plane resolution. Second, we propose an attention module to enable the CNN to focus on the target region. 
Unlike previous works~\cite{Oktay2018}, we explicitly supervise the learning of attention maps so that they can highlight the target structure better. Finally, we propose a voxel-level hardness-weighted Dice loss function to boost the performance of CNNs. The proposed method was validated with T2-weighted MR images of 245 patients with VS tumor. 

\section{Methods}
\begin{figure*}[t]
	\centering 
	\includegraphics[width=1.0\textwidth]{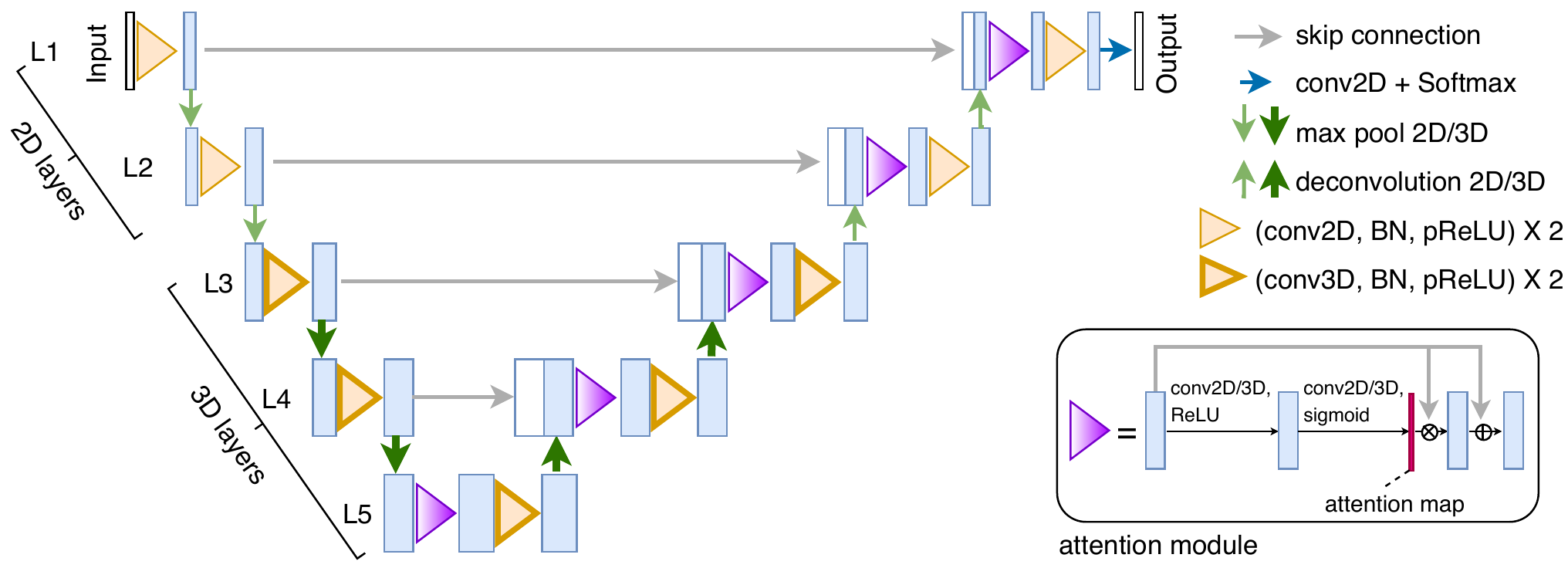}
	\caption{
		The proposed 2.5D U-Net with spatial attention for VS tumor segmentation from anisotropic MRI. The attention module is depicted in the right bottom corner.}
	\label{fig:unet2d5}
\end{figure*}
\subsubsection{2.5D CNN for Segmentation of Images with Anisotropic Resolutions.} 
For our images with high in-plane resolution and low through-plane resolution, 2D CNNs applied slice-by-slice will ignore inter-slice correlation. Isotropic 3D CNNs may need to upsample the image to an isotropic 3D resolution to balance the physical receptive field (in terms of mm rather than voxels) along each axis, which requires more memory and may limit the depth or feature numbers of the CNNs. Therefore, it is desirable to design a 2.5D CNN that can not only use inter-slice features but also be more efficient than 3D CNNs. In addition, to make the receptive field isotropic in terms of physical dimensions, the number of convolution along each axis should be different when dealing with such images. In~\cite{Wang17brats}, a 2.5D CNN was proposed for brain tumor segmentation. However, it was designed for isotropically resampled 3D images and limited by a small physical receptive field along the through-plane axis.

We propose a novel attention-based 2.5D CNN combining 2D and 3D convolutions. As shown in Fig.~\ref{fig:unet2d5}, the main structure follows the typical encoder and decoder design of U-Net~\cite{Hefny2015a}. The encoder contains five levels of convolutions. The first two levels (L1-L2) and the other three levels (L3-L5) use 2D and 3D convolutions/max-poolings, respectively. This is motivated by the fact that the in-plane resolution of our VS tumor images is about 4 times that of the through-plane resolution. After the first two max-pooling layers that downsample the feature maps only in 2D, the feature maps in L3 and the followings have a near-isotropic 3D resolution. At each level, we use a block of layers containing two convolution layers each followed by batch normalization (BN) and parametric rectified linear unit (pReLU). The number of output feature channels at level $l$ is denoted as $N_l$. $N_l$ is set as $16l$ in our experiments. The decoder contains similar blocks of 2D and 3D layers. Additionally, to deal with the small target region, we add a spatial attention module to each level of the decoder, which is depicted in Fig.~\ref{fig:unet2d5} and detailed in the following.

\subsubsection{Multi-Scale Supervised Spatial Attention.} Previous works have shown that spatial attention can be automatically learned in CNNs to enable the network to focus on the target region in a large image context~\cite{Oktay2018}. Building upon these works, we further introduce an explicit supervision on the learning of attention to improve its accuracy.
A spatial attention map can be seen as a single-channel image of attention coefficient $\alpha_i \in [0, 1]$ that is a score of relative importance for each spatial position $i$. As shown in Fig.~\ref{fig:unet2d5}, the proposed attention module consists of two convolution layers. For an input feature map at level $l$ with channel number $N_l$, the first convolution layer reduces the channel number to $N_l/2$ and is followed by ReLU. The second convolution layer further reduces the channel number to 1 and is followed by sigmoid to generate the spatial attention map $\mathcal{A}_l$ at level $l$.  $\mathcal{A}_l$ is multiplied with the input feature map. We also use a residual connection in the attention module, as depicted in Fig.~\ref{fig:unet2d5}.

We propose an attention loss to supervise the learning of spatial attention explicitly during training. Let $G$ denote the multi-channel one-hot ground truth segmentation of an image and $G^f$ denote the single-channel binary foreground mask. For attention map $\mathcal{A}_l$ at level $l$, let $G^f_l$ denote the average-pooled version of $G^f$ so that it has the same resolution as $\mathcal{A}_l$. Our loss function for training is:
\begin{align}
\label{eq:loss}
\mathcal{L} = \frac{1}{L}\sum_l{\ell(\mathcal{A}_l, G^f_l)} + \ell(P,G)
\end{align}
where $L$ is the number of resolution levels ($L=5$ in our case). $\ell(\mathcal{A}_l, G^f_l)$ measures the difference between $\mathcal{A}_l$ and $G^f_l$. It drives the attention maps to be as close to the foreground mask as possible. 
$P$ denotes the prediction output of CNN, i.e., the probability of belonging to each class for each voxel. $\ell(P,G)$ is the segmentation loss.  The multi-scale supervision in Eq.~\eqref{eq:loss} is similar to the holistic loss~\cite{Xie2015}. However, here we apply it to multi-scale attention maps rather than the network's final prediction output. The two terms in Eq.~\eqref{eq:loss} share the same underlying loss function $\ell$, as discussed in the following.

\subsubsection{Voxel-Level Hardness-Weighted Dice Loss.} A good choice of $\ell$ is the Dice loss~\cite{Milletari2016} proposed to train CNNs for binary segmentation, and it has shown good performance in dealing with imbalanced foreground and background classes. For segmentation of small structures with low contrast, some voxels are harder than the others to learn. Treating all the voxels for a certain class equally as in~\cite{Milletari2016} may limit the performance of CNNs on hard voxels. Therefore, we propose automatic hard voxel weighting in the loss function by defining a voxel-level weight:
\begin{align}
\label{eq:w_ci}
w_{ci} = \lambda*abs(p_{ci} - g_{ci}) + (1.0 - \lambda)
\end{align}
where $p_{ci}$ is the probability of being class $c$ for voxel $i$ predicted by a CNN, and $g_{ci}$ is the corresponding ground truth value.  $\lambda \in [0, 1]$ controls the degree of hard voxel weighting. Our proposed  hardness-weighted Dice loss (HDL) is defined as:
\begin{align}
\label{eq:hdl}
\ell_{HDL}(P,G) = 1.0 - \frac{1}{C}\sum_{c}\frac{2\sum_i w_{ci}{p_{ci}g_{ci}} + \epsilon}
{\sum_i w_{ci}({p_{ci} + g_{ci})} + \epsilon}
\end{align}
where $C$ is the channel number of $P$ and $G$, and $\epsilon$=$10^{-5}$ is a small number for numerical stability. 
Similarly to \cite{Milletari2016}, the gradient of $\ell_{HDL}$ with respect to $p_{ci}$ can be easily computed. Note that for the first term $\ell_{HDL}(\mathcal{A}_l, G^f_l)$ in Eq.~\eqref{eq:loss} dealing with attention maps, the channel number is one.

\section{Experiments and Results}
\subsubsection{Data and Implementation.} T2-weighted MRI of 245 patients with a single sporadic VS tumor were acquired in axial view before 
radiosurgery treatment, with high in-plane resolution around 0.4 mm$\times$0.4 mm, in-plane size 512$\times$512, slice thickness and inter-slice spacing 1.5 mm, and slice number 19 to 118. The ground truth was manually annotated by an experienced neurosurgeon and physicist. We randomly split the images into 178, 20 and 47 for training, validation and testing respectively. Each image was cropped with a cubic box of size 100 mm$\times$50 mm$\times$50 mm manually, and normalized by its intensity mean and standard deviation. The CNNs were implemented in Tensorflow and NiftyNet~\cite{Gibson2018} on a Ubuntu desktop with 
an NVIDIA GTX 1080 Ti GPU. For training, we used Adam optimizer with weight decay $10^{-7}$, batch size 2. The learning rate was initialized to $10^{-4}$ and halved every 10k. The training was ended when performance on the validation set stopped to increase. For quantitative evaluation, we measured Dice, average symmetric surface distance (ASSD) and relative volume error (RVE) between segmentation results and the ground truth.

\subsubsection{Comparison of Different Networks.} First, we evaluate the performance of our 2.5D network, and refer to our CNN without the attention module as 2.5D U-Net. Its 2D and 3D counterparts with the same configuration except the dimension of convolution/decovolution and max-pooling are referred to as 2D U-Net and 3D U-Net respectively. For 3D U-Net, the images were resampled to isotropic resolution of 0.4 mm$\times$0.4 mm$\times$0.4 mm. The performance of these networks trained with Dice loss is shown in Table~\ref{tab:eval_1}. It can be observed that our 2.5D U-Net achieves higher accuracy than its 2D and 3D counterparts. In addition, it is more efficient than the other two. Its lower inference time than slice-by-slice 2D U-Net is due to the 3D down-sampled feature maps in L3-L5.
\begin{table*}
	\centering
	\caption{Quantitative evaluation of different networks for VS tumor segmentation. Dice loss was used for training. AG: The attention gate proposed in~\cite{Oktay2018}. PA: Our proposed attention module. SpvPA: The proposed attention with supervision. * denotes significant improvement from 2.5D U-Net based on a paired $t$-test ($p<$ 0.05). }
	\label{tab:eval_1}
	\begin{tabular}
		{p{0.27\linewidth}|
			p{0.16\linewidth} p{0.16\linewidth} p{0.16\linewidth} p{0.16\linewidth} p{0.16\linewidth}|} 
		\hline
		Network 	& Dice(\%) & ASSD (mm) & RVE (\%) & Time (s) \\ \hline
		2D U-Net  &
		80.38$\pm$10.42&
		0.92$\pm$0.68 & 18.01$\pm$17.23 & 3.56$\pm$0.36 \\
		
		3D U-Net  &
		83.61$\pm$13.69&
		0.84$\pm$0.62 & 18.01$\pm$17.48 & 3.90$\pm$0.49 \\
		
		2.5D U-Net &
		85.69$\pm$7.07&
		0.67$\pm$0.45 & 16.02$\pm$14.71 & \textbf{3.49$\pm$0.39}\\ 
		
		2.5D U-Net + AG~\cite{Oktay2018}\ &
		85.93$\pm$6.96&0.58$\pm$0.41* & 15.45$\pm$12.37 & 3.51$\pm$0.34\\ 
		
		2.5D U-Net + PA &
		86.09$\pm$6.94& 0.55$\pm$0.32* & 14.87$\pm$12.19 & 3.52$\pm$0.37\\
		
		2.5D U-Net + SpvPA  &  \bf{86.71$\pm$4.99}*& \textbf{0.53$\pm$0.29}* & \textbf{13.40$\pm$9.34}*  & {3.52$\pm$0.37}
		\\ \hline
		
	\end{tabular}
\end{table*}
\begin{figure*}[t]
	\centering 
	\includegraphics[width=1.0\textwidth]{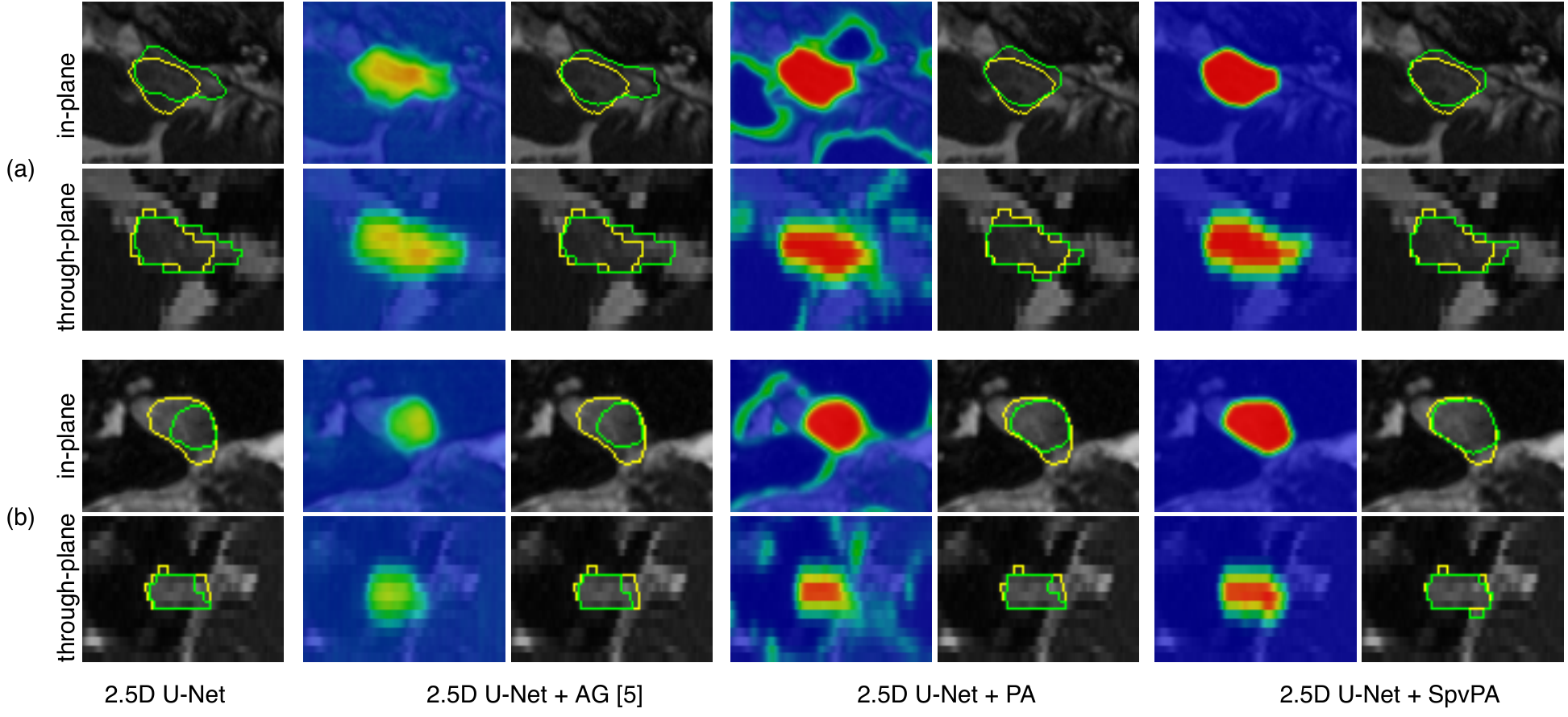}
	\caption{
		Visual comparison of different attention mechanisms for VS tumor segmentation. Odd columns: segmentation results (green curves) and the ground truth (yellow curves). Even columns: attention maps at the highest resolution level (L1) of the decoder, where warmer color represents higher attention.}
	\label{fig:visual_comp}
\end{figure*}
\subsubsection{Effect of Supervised Attention.} 
We further investigated the effect of our proposed attention (PA) module and supervised attention (SpvPA). PA was compared with the attention gate (AG) module proposed in~\cite{Oktay2018}. We combined these modules with our 2.5D U-Net respectively. AG was used to calibrate features from the encoder, as implemented in~\cite{Oktay2018}, while our PA and SpvPA were designed to calibrate the concatenation of encoder and decoder features, as shown in Fig.~\ref{fig:unet2d5}. These variants were trained with Dice loss and their performance is shown in Table~\ref{tab:eval_1}. It can be observed that both AG~\cite{Oktay2018} and PA lead to a better segmentation accuracy than the 2.5D U-Net without attention, and our proposed PA performs slightly better than AG~\cite{Oktay2018}. By using our SpvPA, the segmentation accuracy can be further improved from that of PA. 

Fig.~\ref{fig:visual_comp} shows a visual comparison of these three different attention methods. 
It can be observed that the attention map of AG~\cite{Oktay2018} successfully suppresses most of the background region, but the magnitude for the target region is lower than that of PA and SpvPA. The attention map of PA highlights the target region, but also assigns high attention coefficients for strong edges in the input image. This is mainly because the input for the PA module is a concatenation of high-level and low-level features. Benefiting from our explicit supervision on the learning of attention, the attention map of SpvPA focuses more on the target region and is less blurry than that of AG~\cite{Oktay2018}.  
\begin{figure*}[t]
	\centering 
	\includegraphics[width=1.0\textwidth]{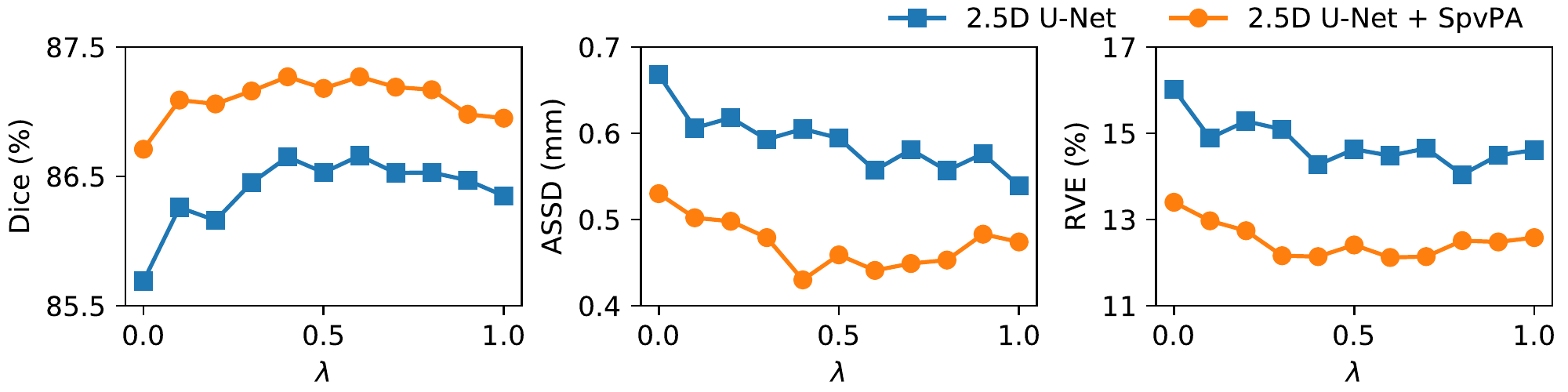}
	\caption{
		Performance of 2.5D U-Net and 2.5D U-Net + SpvPA trained with the proposed voxel-level hardness-weighted Dice loss (HDL) with different values of $\lambda$.}
	\label{fig:plot_focal}
\end{figure*}
\subsubsection{Performance of Voxel-Level Hardness-Weighted Dice Loss.} We additionally used HDL to train 2.5D U-Net and 2.5D U-Net + SpvPA respectively. The average Dice, ASSD and RVE of these two networks with different values of $\lambda$ are shown in Fig.~\ref{fig:plot_focal}. Note that $\lambda = 0.0$ is the baseline without hard voxel weighting and a higher value of $\lambda$ corresponds to assigning higher weights to harder voxels during training. The figure shows that our HDL with different values of $\lambda$ leads to higher segmentation performance. An improvement of accuracy is observed when $\lambda$ increases from 0.0 to 0.4.  Interestingly, when $\lambda$ is higher than 0.6, the segmentation accuracy decreases, as shown by the curves of Dice in Fig.~\ref{fig:plot_focal}. This indicates that giving too much emphasis to hard voxels may decrease the generalization ability of the CNNs. As a result, we suggest a proper range of $\lambda$ as [0.4, 0.6]. Quantitative comparison between Dice loss~\cite{Milletari2016} and our proposed HDL with $\lambda = 0.6$ is presented in Table~\ref{tab:eval_2}. It shows that our proposed HDL outperforms Dice loss~\cite{Milletari2016} for both 2.5D U-Net and 2.5D U-Net + SpvPA.
\begin{table*}
	\centering
	\caption{Performance of two CNNs trained with different loss functions. * denotes significant improvement from Dice loss~\cite{Milletari2016} based on a paired $t$-test ($p < $ 0.05).}
	\label{tab:eval_2}
	\begin{tabular}
		{p{0.26\linewidth}|p{0.18\linewidth}|
			p{0.16\linewidth} p{0.16\linewidth} p{0.16\linewidth} p{0.16\linewidth} p{0.16\linewidth}} 
		\hline
		Network & Training loss 	& Dice(\%) & ASSD (mm) & RVE (\%) \\ \hline
		\multirow{2}{*}{2.5D U-Net}  & Dice loss~\cite{Milletari2016}  &
		85.69$\pm$7.07 & 0.67$\pm$0.45 & 16.02$\pm$14.71 \\
		& HDL ($\lambda$=0.6) &
		86.66$\pm$6.01*&
		0.56$\pm$0.37* & 14.48$\pm$12.30* 
		\\ \hline
		\multirow{2}{*}{2.5D U-Net + SpvPA } & Dice loss~\cite{Milletari2016}  &
		86.71$\pm$4.99 & 0.53$\pm$0.29 & 13.40$\pm$9.34 \\
		& HDL ($\lambda$=0.6) &
	   \bf{87.27$\pm$4.91} &
		\bf{0.43$\pm$0.31*} & \bf{12.14$\pm$8.94} \\
	    \hline
		
	\end{tabular}
\end{table*}
\section{Discussion and Conclusion}
In this work, we propose a 2.5D CNN for automatic VS tumor segmentation from high-resolution T2-weighted MRI. Our network is a trade-off between standard 2D and 3D CNNs and specifically designed for images with high in-plane resolution and low through-plane resolution. Experiments show that it outperforms its 2D and 3D counterparts in terms of segmentation accuracy and efficiency. To deal with the small target region, we propose a multi-scale spatial attention mechanism with explicit supervision on the learning of attention maps. Experimental results demonstrate that the supervised attention can guide the network to focus more accurately on the target region, leading to higher accuracy of the final segmentation. We also combine automatic hard voxel weighting with existing Dice loss~\cite{Milletari2016}, and the proposed voxel-level hardness-weighted Dice loss can lead to further performance improvement. 
This will facilitate the rapid adoption of these techniques into clinical practice, providing clinicians with the means for accurate automatically-generated segmentations that will be used to inform patient management decisions. Though our methods can also be applied to ceT1 images, this work on T2 image segmentation improves patient safety by enabling patients to undergo serial imaging without the need to use  potentially harmful contrast agents.

\subsubsection{Acknowledgements.}\label{sec:acknowledgements}
This work was supported by Wellcome Trust [203145Z/16/Z; 203148/Z/16/Z; WT106882], and EPSRC [NS/A000050/1; NS/A000049/1] funding. TV is supported by a Medtronic / Royal Academy of Engineering Research Chair [RCSRF1819/7/34].
%
%
\bibliographystyle{splncs03}
\bibliography{./miccai2019}


%
%

\end{document}